\documentclass{aa} \usepackage{graphicx}
\newcommand{\oversim}[2]{\protect{\mbox{\lower0.5ex\vbox{%
   \baselineskip=0pt\lineskip=0.2ex
   \ialign{$\mathsurround=0pt #1\hfil##\hfil$\crcr#2\crcr\sim\crcr}}}}} 
\newcommand{\simgreat}{\mbox{$\,\mathrel{\mathpalette\oversim>}\,$}} 
\newcommand{\simless} {\mbox{$\,\mathrel{\mathpalette\oversim<}\,$}} 
\usepackage{txfonts}
\begin{document}
\title{The great disk of Milky-Way satellites and cosmological
  sub-structures}

    \author{Pavel Kroupa{\inst{1,}\inst{2,}}\thanks{\it Heisenberg Fellow} \and
            Christian Theis\inst{1,}\inst{3} \and
            Christian M. Boily\inst{4}
           }

    \offprints{P. Kroupa,\\
               \email{pavel@astro.uni-bonn.de}
               }

    \institute{Institut f\"ur Theoretische Physik und Astrophysik der
                   Universit\"at Kiel, D--24098 Kiel, Germany
          \and
               Sternwarte Bonn, Auf dem H\"ugel 71, D--53121 Bonn, Germany
          \and
               Institut f\"ur Astronomie der Univ.\ Wien,
                  T\"urkenschanzstr.\ 17, A--1180 Vienna, Austria\\
                  \email{theis@astro.univie.ac.at}
          \and
              Observatoire Astronomique de Strasbourg, 11, rue de
                  l'Universit\'e, F--67000 Strasbourg, France\\
                  \email{cmb@pleiades.u-strasbg.fr}
              }

    \date{Received; accepted}
    
    \abstract{We show that the shape of the observed distribution of
      Milky Way (MW) satellites is inconsistent with being drawn from
      a cosmological sub-structure population with a confidence of
      99.5 per cent. Most of the MW satellites therefore cannot be
      related to dark-matter dominated satellites. 
      }

    \maketitle
%
\section{Introduction}
\label{sec:intro}

Calculations of structure formation within the framework of cold dark
matter (CDM) cosmology show that Milky Way (MW) type systems have the
same scaled theoretical distribution of sub-haloes as rich galaxy
clusters, and within 500~kpc they should contain about 500 sub-haloes
with masses $M\simgreat 10^8\,M_\odot$ (Moore et al.\ \cite{moore99};
Klypin et al.\ \cite{klypin99}; Governato et al. 2004). However, only
13~dwarves have been found within a distance of 500~kpc around the MW.
The observed dwarves may only sample a sub-set of the actually present
CDM sub-structures (Stoehr et al. 2002; Hayashi et al. 2003; Bullock
et al. 2000; Susa \& Umemura 2004; Kravtsov, Gnedin \& Klypin 2004).
Such biasing could be the result of complex early baryonic physics
that cannot, at present, be treated theoretically in sufficient
detail, but Kazantzidis et al. (2004) point out that this cannot be
the entire solution.

An additional path to testing predictions of CDM cosmology is to
compare the shape of the observed satellite distribution to the
theoretical shapes (Zaritsky \& Gonzalez 1999; Hartwick 2000; Sales \&
Lambas 2004).  The sub-structures fall inwards from filaments that are
thicker than the virialised regions of the hosts. However, within its
virialised region, the number distribution of sub-structure in a
theoretical host halo follows that of its dark-matter (DM)
distribution.  CDM models predict the host DM haloes to be oblate with
flattening increasing with increasing mass and radius (Combes 2002;
Merrifield 2002). The ratio of minor to major axis of the DM density
distribution has the value $q_{\rm d}=0.7\pm0.17$ for MW-sized haloes
within the virial radius.  The intermediate to major axis ratio is
$q_{\rm d}'\simgreat 0.7$ (Bullock 2002).  When dissipational baryonic
physics is taken into account the haloes become more axis-symmetric
(larger $q_{\rm d}'$) and more flattened, $q_{\rm d}=0.5\pm0.15$
within the virial radius.  The minor axis is co-linear to the angular
momentum of the baryonic disk (Dubinski 1994). Prolate haloes do not
emerge.  The empirical evidence is that the MW dark halo is somewhat
flattened (oblate) with $q_{\rm d}\simgreat 0.8$ within $R\simless
60$~kpc (Olling \& Merrifield 2000, 2001; Ibata et al.  2001; Majewski
et al.  2003; Mart{\'{\i}}nez-Delgado et al. 2004).  Beyond this
distance the shape is likely to be more oblate (Bullock 2002), but
invoking continuity the axis ratio $q_{\rm d}$ cannot change
drastically.  The theoretical sub-structure distribution of MW-type
hosts must therefore be essentially isotropic (Ghigna et al.
  1998; Zentner \& Bullock \cite{zentner03}; Diemand, Moore \& Stadel
2004; Kravtsov et al. 2004; Aubert et al.\ \cite{aubert04}).

If the MW dwarves do indeed constitute the shining fraction of DM
sub-structures, then their number-density distribution should be
consistent with an isotropic (i.e.  spherical) or oblate power-law
radial parent distribution.  This is assumed to be the case by most
researchers, given the relatively small number of satellites. With
this contribution we show that, despite its smallness, the MW
satellite sample is inconsistent with a cosmological sub-structure
population. We do this by concentrating on the most elementary facts,
namely purely on the positions of the satellites.

\section{Dwarf galaxies near the Milky Way}
\label{sec:MWsat}

Table~\ref{tab:MWsat} lists distances and coordinates of the $N=16$
dwarves closest to the MW. Given these data Galactocentric coordinates
are calculated, $X_{\rm D} = -D_\odot + D\,{\rm sin}(90^{\rm
o}-b)\,{\rm cos}(l), Y_{\rm D} = D\,{\rm sin}(90^{\rm o}-b)\,{\rm
sin}(l), Z_{\rm D} = D\,{\rm cos}(90^{\rm o}-b)$, with uncertainties
derived from the uncertainties in $D$.
\begin{table*}
\begin{center}
      \caption{Dwarf galaxies
        within the vicinity of the MW. The first column is a running
        number used throughout this text; the parentheses contain the
        running dwarf number used in \S~\ref{sec:prob} after excluding
        SMC and UMi. $D$ and $eD$ are the distance and its
        uncertainty, respectively. $l$, $b$ are the galactic longitude
        and latitude, respectively, as seen from the Sun and defined
        such that $l=0, b=0$ points towards the Galactic centre which
        is assumed to lie at a distance $D_\odot=8.5$~kpc from the
        Sun, and $l$ increases in anticlockwise direction.  The
        Galactocentric distance of the dwarf is given by $R$.  The
        name of the dwarf is contained in the 7th column.  The data
        are from Mateo (\cite{mateo98}, table~2), except that for the
        LMC $D$ and $eD$ are taken from Salaris et al.\ 
        (\cite{salaris03}) and Clementini et al.\ 
        (\cite{clementini03}), and likewise for the SMC from Dolphin
        et al.\ (\cite{dolphin01}).  The remaining columns contain the
        plane-fitting results for the innermost $N$ dwarfs
        (\S~\ref{sec:plane}): $R_{\rm cut}$ is the largest distance to
        the Galactic centre of this sample, and the fitted plane has a
        root-mean square height $\Delta$ and a distance to the
        Galactic centre $D_{\rm P}$. For comparison, the final column
        lists the root-mean square height $\Delta_2(R_{\rm cut})$ for
        samples of $4\times10^5$ theoretical dwarves with an isotropic
        isothermal radial number density profile ($p=2$) and radial
        cutoff $R_{\rm cut}$.
        \label{tab:MWsat}}

    \vspace*{0.2cm}
    \begin{tabular}[h!]{ccccccccccccc} \hline

      satellite &$D$     &$eD$    &$l$     &$b$   &$R$ 
     &Name &$R_{\rm cut}$ &$\Delta$ &$\Delta/R_{\rm cut}$
       &$D_{\rm P}$ &&$\Delta_2$ \\

      number    &[kpc]   &[kpc]   &[deg]   &[deg] &[kpc] 
     & & [kpc] &[kpc]    & &[kpc]     & &[kpc]\\

           \hline

      1(1)     &  24     &   2      &   5.6  & $-14.1$ &  16  
     & Sgr &-- &-- &-- &-- &&--\\
      2(2)     &  50.8   &   2.2    & 280.5  & $-32.9$ &  50  
     & LMC &-- &-- &-- &-- &&--\\
      3(-)     &  59.7   &   2.2    & 302.8  & $-44.3$ &  57  
     & SMC &-- &-- &-- &-- &&--\\
      4(-)     &  66     &   3      & 105.0  & $+44.8$ &  68  
     & UMi &  68 &   2.5 &0.04 & 0.6 &&  23 \\
      5(3)     &  79     &   4      & 287.5  & $-83.2$ &  79  
     & Sculptor&  79 &  11.8 &0.15 & 3.0 &&  27\\
      6(4)     &  82     &   6      &  86.4  & $+34.7$ &  82  
     & Draco   &  82 &  11.0 &0.13 & 3.2 &&  28\\
      7(5)     &  86     &   4      & 243.5  & $+42.3$ &  89  
     & Sextans &  89 &  13.5 &0.15 & 1.2 &&  30\\
      8(6)     & 101     &   5      & 260.1  & $-22.2$ & 103  
     & Carina  & 103 &  14.2 &0.14 & 1.4 &&  34\\
      9(7)     & 138     &   8      & 237.1  & $-65.7$ & 140  
     & Fornax  & 140 &  23.9 &0.17 & 2.0 &&  47\\
     10(8)     & 205     &  12      & 220.2  & $+67.2$ & 208  
     & LeoII   & 208 &  23.2 &0.11 & 1.9 &&  69\\
     11(9)     & 250     &  30      & 226.0  & $+49.1$ & 254  
     & LeoI    & 254 &  26.4 &0.10 & 1.9 &&  85\\
     12(10)    & 445     &  30      & 272.2  & $-68.9$ & 445  
     & Phoenix & 445 &  32.0 &0.07 & 2.1 && 148\\
     13(11)    & 490     &  40      &  25.3  & $-18.4$ & 483  
     & NGC 6822& 483 &  86.4 &0.18 & 3.5 && 161\\
     14(12)    & 690     & 100      & 196.9  & $+52.4$ & 695  
     & Leo A   & 695 & 100   &0.14 & 3.5 && 232\\
     15(13)    & 880     &  40      & 322.9  & $-47.4$ & 875  
     & Tucana  & 875 & 123   &0.14 & 3.5 && 292\\
     16(14)    & 955     &  50      &  94.8  & $-43.5$ & 956  
     & Pegasus & 956 & 159   &0.17 & 3.3 && 319\\

\hline

    \end{tabular}
\end{center}
\end{table*}
The data are compared with an isotropic power-law density
distribution, $\rho(R) = \rho_o\,R^{-p}$, where $R=\sqrt{X_{\rm D}^2 +
  Y_{\rm D}^2 + Z_{\rm D}^2}$. Using the Kolmogorov-Smirnov (KS) test
it can be shown that the cumulative dwarf sample is consistent with a
radial near-isothermal density distribution, the 5~per cent confidence
margin being $1.8<p<2.6$ for the $N=11$ innermost dwarves.  The
solutions shift to larger $p$ as further outlying dwarves are added.
This is a similar behaviour as seen in theoretical distributions of
sub-haloes (e.g.\ fig.~5 in Zentner \& Bullock \cite{zentner03}).

\begin{figure}
     \resizebox{\hsize}{!}{\includegraphics{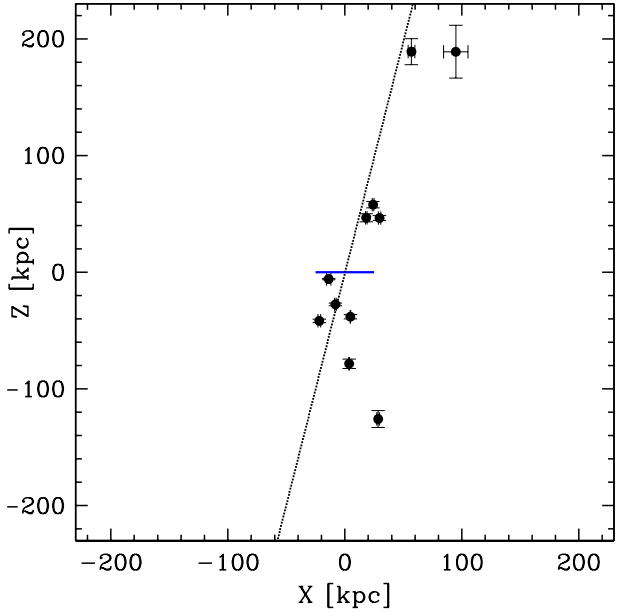}}
     \caption
     {The position of the innermost~11 MW satellites
       (Table~\ref{tab:MWsat}) as viewed from a point located at
       infinity and $l=167^{\rm o}.91$.  The MW disk is indicated by
       the horizontal line $-25 \le X/{\rm pc} \le 25$, and the centre
       of the coordinate system lies at the Galactic centre. The
       dashed line marks the fitted plane for $N=11$ seen edge-on in
       this projection.}
      \label{fig:pos2}
\end{figure}

\begin{figure}
     \resizebox{\hsize}{!}{\includegraphics{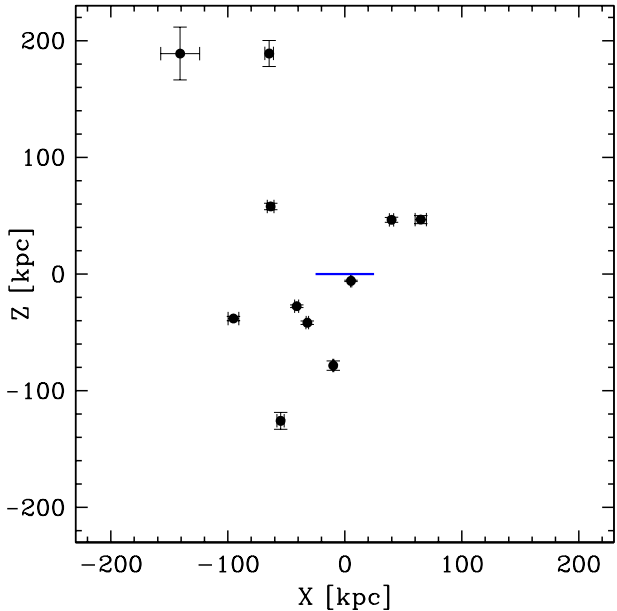}}
     \caption
     {As fig.~\ref{fig:pos2} but viewed from $l=77^{\rm o}.91$. The
       fitted plane is here seen face-on.}
      \label{fig:pos1}
\end{figure}

The data are plotted in Fig.~\ref{fig:pos2} after clockwise rotation
by an angle $\phi=167^{\rm o}.9$ about the Z-axis, $X = X_{\rm D} \,
{\rm cos}\phi + Y_{\rm D}\,{\rm sin}\phi, Y = -X_{\rm D} \, {\rm
  sin}\phi + Y_{\rm D}\,{\rm cos}\phi$ and likewise for the
uncertainties.  The distribution is highly anisotropic and planar in
form. It is the aim of this contribution to quantify the significance
of this anisotropy. A rotation of Fig.~\ref{fig:pos2} by $90^{\rm o}$
shows the distribution to be approximately disk-like
(Fig.~\ref{fig:pos1}).

\section{The satellite plane}
\label{sec:plane}

A plane can be described by the {\sc Hesse} form, $\vec{n} \bullet
\left(\vec{x} - \vec{p}\right) = 0$, where $\vec{n}$ is the normal
vector, $\vec{p}$ a vector pointing from the origin (the centre of the
MW) to a point in the plane, and $\vec{x}$ an arbitrary vector from
the origin to the plane.  With $\vec{n}=(n_1,n_2,n_3)$ and
$\vec{x}(i)=(X(i),Y(i),Z(i))$ being the coordinates of the galaxies,
$d(i) = n_1\,X(i) + n_2\,Y(i) + n_3\,Z(i) - D_{\rm P}$, becomes
identical to Hesse form if $d(i)=0$; $d(i)$ being the distance of the
$i$th dwarf to the plane.  $D_{\rm P}=\vec{n}\bullet\vec{p}$ is the
shortest distance of the plane to the origin. The problem of finding
the plane can thus be reduced to a least-squares linear regression
problem, where the aim is to find the coefficients, $n_i, D_{\rm P}$
with the condition $\sum_{i=1}^3\,n_i^2=1$, that minimise
$\sum_{i=1}^N d^2(i)$. To achieve this the method of normal equations
using Gauss-Jordan elimination is employed to solve the set of linear
equations (Press et al.\ \cite{press92}). For each fitted plane the
root-mean square height of the resulting disk distribution is
calculated, $\Delta(R_{\rm cut}) = \sqrt{(1/N)\,\sum_{i=1}^N d^2(i)}$.
Note that the applied minimisation does not include the location of
the Galactic centre as a constraint.  Thus, in principle the fitted
plane to a small number of dwarves ($N\simless12$) could lie far away
from the Galactic centre. The weights that do enter the regression are
merely given by the uncertainties in distance.  The direction of the
normal vector, or the location of the pole of the plane, $l_{\rm P},
b_{\rm P}$, follows from $\theta = {\rm arccos}(n_3), b_{\rm
  P}'=90^{\rm o}-\theta, l_{\rm P}'={\rm arcsin}((n_2) / {\rm
  sin}(\theta))$. As no kinematical information is included the
direction of the pole is ambiguous, $b_{\rm P}=b_{\rm P}', l_{\rm
  P}=l_{\rm P}'$, or $b_{\rm P}=-b_{\rm P}', l_{\rm P}=l_{\rm
  P}'+180^{\rm o}$.

Table~\ref{tab:MWsat} lists some results of the fitted plane for a
decreasing number of dwarves.  The empirical disk height, $\Delta$, is
always much smaller than the theoretical height, $\Delta_2$, for an
isothermal and isotropic model number density distribution centred on
the origin of the MW.  {\it The MW dwarves thus appear to be
  distributed as a great disk with a ratio of height to radius
  $\simless0.15$}.

\begin{figure}
     \resizebox{\hsize}{!}{\includegraphics{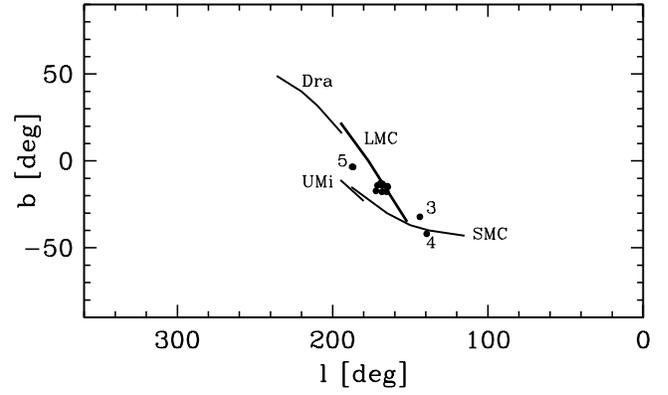}}
     \caption
        {The position on the Galactic sky of the poles of the planes fitted to
         the dwarves of Table~\ref{tab:MWsat}.  Plotted are $b_{\rm P}=-b_{\rm
         P}'$ and $l_{\rm P}=l_{\rm P}'+180^{\rm o}$ and the number of dwarves
         used for the fit ranges from $N=16$ down to $N=3$
         (Table~\ref{tab:MWsat}). The cases for $N=3,4,5$ are indicated with
         numbers. The others cluster very tightly around $l_{\rm
         P}\approx168^{\rm o}, b_{\rm P}\approx-16^{\rm o}$. The likely
         position of the orbital poles of the LMC, SMC, Draco and UMi are
         indicated by the solid curves (from fig.~3 in Palma et al.\
         \cite{palma02}).}
     \label{fig:poles}
\end{figure} 

The poles of the planes and the orbital poles of the dwarves LMC, SMC,
Draco and UMi agree remarkably well (Fig.~\ref{fig:poles}).  {\it This
  is surprising because the results are obtained using completely
  different methods}. The position of the poles of the planes found
here depend only on the spatial distribution of the dwarves. In
contrast, an orbital pole is the direction of the orbital angular
momentum and relies on the direction of the measured proper motion of
the respective object.  Sgr is on a polar orbit but has a kinematical
pole ($l\approx280, b\approx 0$, Palma, Majewski \& Johnston\ 
\cite{palma02}) lying approximately at a right angle to the great disk
{\it and} to the MW disk.  On the basis of the weakly bound core of
Sgr which makes it difficult for Sgr to survive the many orbits
implied by its current angular momentum, Zhao (\cite{zhao98}) proposed
that it may have been scattered into its present low-pericenter orbit
by an encounter with the LMC/SMC about 2--3~Gyr ago.  Sgr contributes
the most deviant cos$(\omega)$ value in the sample because it is
closest to the MW centre and thus high above the local great disk.
Taking Sgr out of the sample would increase the discrepancy,
quantified in \S~\ref{sec:prob}, between the dwarf sample and the
hypothesis that they are the visible cosmological sub-halo population.


\section{The likelihood}
\label{sec:prob}

The null hypothesis is that the $N$ observed dwarves are drawn from a
cosmological population.  We therefore need to establish the
probability that the observed distribution is drawn from a spherical
parent distribution.  

\begin{figure}
     \resizebox{\hsize}{!}{\includegraphics{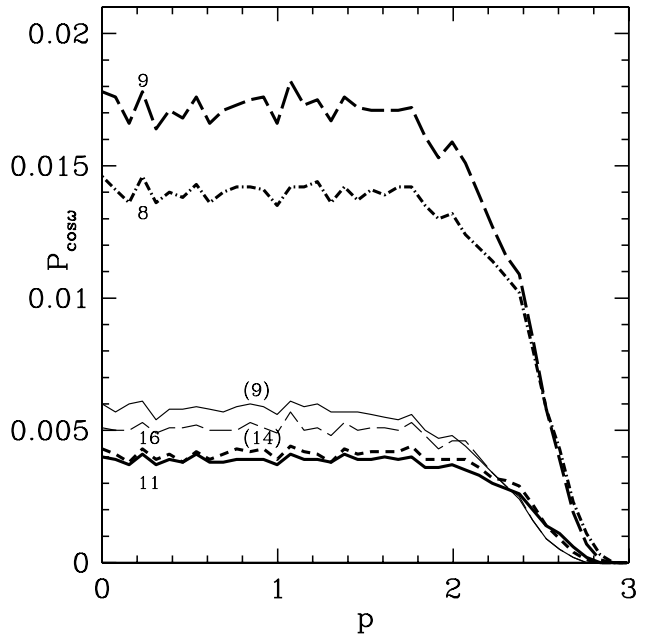}}
     \vspace{-25mm}
     \caption
     {The probability, $P_{{\rm cos}\omega}$, that the observed dwarf
       sample stems from a parent isotropic radial power-law density
       distribution with index $p$. The number of innermost dwarves in
       the sample (Table~\ref{tab:MWsat}) is indicated by the numbers.
       Thus 11, for example, means that the innermost $N=11$ dwarves
       (out to and including LeoI) are compared with the isotropic
       power-law distribution. The thin curves (and numbers in
       parentheses) are probabilities calculated by excluding SMC and
       UMi from the data; here $N=(9)$ incorporates all dwarves except
       SMC and UMi out to and including LeoI (Table~\ref{tab:MWsat}).
       The increase of $P_{{\rm cos}\omega}$ with decreasing $N$ and
       $(N)$ is a result of weakening confidence as the number of
       observed data is reduced.  The decrease of $P_{{\rm
           cos}\omega}$ for $p>2.4$ comes from the theoretical
       distribution being increasingly concentrated towards the
       Galactic centre while the plane lies off-centre ($D_{\rm
         P}>0$).}
     \label{fig:pcosw}
\end{figure}

The vector pointing from the Galactic centre to the closest point,
P$_{\rm cl}$, on the plane is $\vec{d}_{\rm P}=D_{\rm P}\vec{n}$, and
the vector from this point P$_{\rm cl}$ to a dwarf is $\vec{x}' =
-\vec{d}_{\rm P} + \vec{x}$. The angle, $\omega$, between the normal
vector and the dwarf as viewed from P$_{\rm cl}$ is then given by $
{\rm cos}(\omega) = \vec{n} \bullet \vec{x}'/|\vec{x}'|$.  The
cumulative distribution of cos$(\omega)$ about the fitted plane is
calculated for the observed sample using the innermost $N$ dwarves,
and also for $N_{\rm m}=10^5$ model dwarves distributed according to
the theoretical parent radial power-law distribution which is centred
on and isotropic about the Galactic centre.  The KS test quantifies
the confidence that can be placed in the hypothesis that the observed
sample stems from this parent distribution. The results, plotted in
Fig.~\ref{fig:pcosw}, show that this hypothesis can be rejected with a
confidence of better than 98~per cent, and even 99.6~per cent for
$N\ge11$.  This comes about because the real sample is deficient near
the poles of the great disk.

Orbital pole analyses have been showing that the SMC, UMi and the LMC
form a kinematical family (Palma et al.  2002).  Taking these two
objects out of the sample, kinematically-linked dwarves are removed
with the expectation that the remaining dwarves should be more
consistent with the isotropic parent distribution.  As the thin curves
in Fig.~\ref{fig:pcosw} show this is not the case.  Instead, the
probabilities that the $N=(9)$ sample without the SMC and UMi stems
from an isotropic parent distribution is reduced (as compared to the
$N=9$ sample). This comes about because the two dwarves are relatively
close to the Galactic centre thus adding relatively large $\omega$
angles when they are included.

The disk-like distribution of the dwarves lying nearby to the MW noted
in Fig.~\ref{fig:pos2} is therefore highly significant.  The local
dwarves do not stem from an isotropic distribution. {\it Their
  distribution is therefore severely at odds with the sphericity of
  the MW dark matter halo, and even more at odds with an oblate halo
  having the same orientation as the MW disk}.

\section{Concluding remarks}
\label{sec:concs}

Cosmological models can be tested among other ways by comparing the
theoretical sub-structure distribution with observed satellite
distributions. The theoretical distribution contains about 500
sub-haloes within approximately 500~kpc of a MW-type galaxy and
follows an approximately power-law radial distribution with $p\simless
2$ and is essentially isotropic. The well-known MW distribution
contains only a dozen dwarves, is indeed consistent with the
theoretical radial distribution but is highly anisotropic. The
anisotropy is such that the MW dwarves form a disk-like structure with
a root-mean-square height of $10-30$~kpc which lies nearly
perpendicularly to the plane of the MW. The pole of this great disk
lies close to the orbital poles of the LMC, the SMC, Draco and Ursa
Minor. The distance of closest approach of the plane to the Galactic
centre, $D_{\rm P}\la 2$~kpc, is much smaller than the radial extent
of the Galactic disk ($\approx 20$~kpc) or even the root-mean square
height, $\Delta$, of the disk of satellites ($D_{\rm P}\ll \Delta$).
This is a strong indication that the sample of dwarves within about
250~kpc is relaxed in the Galactic potential.  Their orbits must be
confined within the great disk because the likelihood of obtaining
such a disk-like dwarf distribution given a true underlying isotropic
distribution (that ought to match the sphericity of the MW DM halo) is
less than 0.5~per cent. This result persists even after removing the
kinematically related SMC and UMi from the analysis.  A distribution
of polar orbits with arbitrary eccentricities and orientation of
orbital planes is also excluded with the same confidence because it
leads to an isotropic distribution of dwarves.  An oblate MW dark
matter halo would yield an even larger discrepancy with the disk of
satellites.

An alternative approach is taken by Hartwick (2000) who argues that
the 10~satellites within 400~kpc (the LMC and SMC are combined to one
satellite) map the MW DM halo shape and form a highly inclined and
highly prolate system with minor/major axis ratio $q_{\rm d}\approx
0.03-0.05$.  However, the extreme triaxiality derived in this way is
completely inconsistent with the observational and theoretical shapes
of CDM host-haloes and sub-structure distributions
(\S~\ref{sec:intro}).
  
The approach taken here differs by noting the very significant
mismatch between (i)~the disk-like satellite distribution, (ii)~the
{\it independent}\, empirical constraints on the shape of the MW dark
matter halo, and (iii)~the theoretical shapes of CDM host haloes
(\S~\ref{sec:intro}). In the view presented here, the mismatch between
the number {\it and}\, spatial distribution of MW dwarves compared to
the theoretical distribution challenges the claim that the MW dwarves
are cosmological sub-structures that ought to populate the MW halo.

A more natural and more conservative (by not resorting to exotic
physics) explanation for the MW dwarf distribution in a great disk
with a ratio of height to radius of~0.1--0.2 would appear to be in
terms of a causal connection between most of them.  This could be the
case if most of the dwarves stem from one initial gas-rich parent
satellite on an eccentric near-polar orbit that interacted with the
young MW, perhaps a number of times, forming tidal arms
semi-periodically as its orbit shrank. The early gas-rich tidal arms
may have condensed in regions to tidal dwarf galaxies, as is observed
in present-day interacting gas-rich galaxies (e.g. Knierman et al.
2003; Weilbacher, Duc \& Fritze-v.~Alvensleben 2003).  The LMC may be
the most massive remnant of this larger satellite, while the lesser
dwarves may be its old children (Lynden-Bell \cite{lyndenbell76}).
The Magellanic Stream may be just such a newly formed but meagre tidal
feature (Kunkel \cite{kunkel79}), and the alignment of the disk of
satellites with the surrounding matter distribution (Hartwick 2000)
may simply result from the gas-rich parent satellite coming-in from
that direction.  The different chemical enrichment and star-formation
histories of the various dwarves (e.g.\ Ikuta \& Arimoto
\cite{ikuta02}; Grebel et al.\ \cite{grebel03}) may in this case be a
result of their different initial masses that will have been
significantly larger than their present-day baryonic masses (Kroupa
\cite{kroupa97}) and the complex interplay between stellar evolution,
tides, gaseous stripping and gas accretion during the orbits within
the MW halo, none of which are presently understood in much detail.
The simulations of Kroupa have shown that ancient tidal-dwarf galaxies
may appear similar to some of the observed dSph satellites.

The sub-structure under-abundance problem extends to fossil galaxy
groups where early photo-evaporation could not have removed baryons
from the sub-structures (D'Onghia \& Lake 2004), and a
sub-structure-overabundance is evident for rich clusters (Diemand et
al. 2004). CDM cosmology thus faces a sub-structure challenge on all
mass scales.


\end{document}